# Effect of administration gamma-amino butyric acid on physiological performance of broiler chicks

Hasanain N. Ezzat,[*,1] Ihsan M. Shihab [†]

*Department of Animal Production, Faculty of Agricultural Engineering Sciences, Baghdad University, Iraq; [†] and Department of Pathology and Poultry Diseases, College of Veterinary Medicine, University of Baghdad-Iraq.

## ABSTRACT

This experiment was conducted to study the effect of administration γ-amino butyric acid (GABA) on physiology performance of broiler chicks. Ninty six (Ross 308) of 1-35 day old chicks with an initial weight of 38g and distributed randomly among four treatment groups with four replicates by six chicks for each replicate. The following treatments were used: T1:control treatment, T2: the birds were administration 0.2ml 0.4% GABA solution daily, T3: the birds were administration 0.2ml 0.5% GABA solution daily and T4: the birds were administration 0.2ml 0.6% GABA solution daily.

The results of the experiment indicated that there were no significant differences for the high-density lipoprotein and triglyceride trait when calculating blood lipids in the third week, while administration-level treatments recorded a significant decrease when calculating low-density lipoprotein and cholesterol, and there were no significant differences between the treatments at the fifth week of the experiment. When calculating blood enzymes, there are significant differences between the treatments at the third week, where the second and third treatment decreased significantly when calculating AST and ALT, and in the fifth week, the third treatment recorded a significant decrease when calculating ALT, while there were no significant differences when calculating AST. The results also indicated that there were no significant differences between the treatments in the third and fifth week of the experiment when calculating the total protein and the concentration of glucose, while significant differences were observed between the treatments in the fifth week of the experiment when measuring the concentration of calcium and phosphorus where administration treatments recorded a significant increase while there were no significant differences between the treatments In the third week. A significant decrease in administration treatments was observed when calculating the concentration of uric acid in the third and fifth week of the experiment.

---

[1] Corresponding author . E-mail: hasanain.nashat@gmail.com

## INTRODUCTION

Gamma aminobutyric acid (GABA) is a non-essential amino acid and is a non-protein amino acid (Park and kim 2015 and Li et al. 2010). The chemical formula is C4 H9 NO2 and represents about 30% of the neurotransmitters in the central nervous system that is generally found in nature as it is found in the brain, heart and kidneys as well as found in fetuses, rice in plants and in bacteria and yeast (Hashimoto,2011), GABA is one of the primary inhibitors in the central nervous system (Jonaidi et al. 2012) that contribute is to reduce stress intensity ( Chen et al. 2014 and Wang et al. 2011) , In addition to its function of the neurotransmission, it performs a variety of other physiological functions, such as enhancing memory and enhancing immunity against stress, and it is also sedative and low blood pressure (Hayakawa et. al 2004). The lack of gamma amino butyric acid in the body leads to the emergence of symptoms that similar to epilepsy and is considered as anti-depressants in humans and works to remove tension (Omori et al, 1987; Oh and Choi, 2000; and Jin, 2013). These studies have shown the possibility of producing (gamma) by fermentation using bacteria, fungi and yeast (Smith et al, 1992 and Lu et al, 2008).

Studies indicated that gamma amino butyric acid has an important role to reduce the heat production as well as the fever (Shekhar et al, 2006) and has positive effects on the growth performance of the broiler ( Dai at el. 2011), as well as the immune activity and oxidative function in chickens subject to heat stress ( Zhang et al, 2012 and Chen et al, 2013,2014).

The present study aimed to investigate the effect of administration on improving the physiological performance of broiler chickens

## Materials and methods

This study was carried out in the poultry section of the animal production program in the College of Agricultural Engineering Sciences, the University of Baghdad. In this study 96 (Ross 308) of 1-35 day old chicks with an initial weight of 38g are distributed randomly among four treatment groups with four replicates by six chicks for each replicate. The chicks were fed with a starter diet from day 1-21 and with a finisher diet from day 21-35 as shown in Table 1. The following treatments were used: T1:control treatment, T2: the birds were administration0.2ml 0.4% GABA solution daily, T3: the birds were administration0.2ml 0.5% GABA solution daily and T4: the birds were administration 0.2ml 0.6% GABA solution daily.

Table 1. Proportion of feed material used in feed and its chemical composition

| Components | Starter diet % | Finisher diet % |
|---|---|---|
| Corn | 30.5 | 39 |
| Wheat | 29 | 24.2 |
| Soybean | 30.5 | 25.3 |
| Protein Concentrated | 5 | 5 |
| Sun flower Oil | 2.9 | 4.2 |
| D.C.P | 0.7 | 0.9 |
| Limestone | 0.9 | 0.9 |
| NaCl | 0.3 | 0.3 |
| Minerals and Vitamin mixture | 0.2 | 0.2 |
| Total | 100 | 100 |
| *chemical compositions | | |
| Protein (%) | 22.82 | 20.49 |
| Metabolism Energy(Kilogram/Kilocalorie) | 3041.75 | 3166.86 |
| Ca % | 0.84 | 0.80 |
| P % | 0.46 | 0.50 |
| Meth % | 0.50 | 0.47 |
| Meth + Cy % | 0.86 | 0.77 |

*The chemical composition of feeding in the diet was estimated according to NRC1994

*Blood tests*

Blood samples were collected randomly for five birds of 3 and 5 weeks age from each treatment by stitching the wing from the brachial vein and collecting blood in test tubes that did not contain anticoagulant and then placed them horizontally. The samples were then placed in the centrifuge at a speed of 3000 cycles/minute for 15 minutes to separate the plasma by other sealed tubes and frozen under -15 (-20) until laboratory tests. Many Variables were measured such as blood protein, blood lipids, blood enzymes, glucose, phosphorus, calcium, and uric acid.

*Statistical analysis*

Complete Randomized Design (CRD) was used to investigate the effect of the studied treatments on different traits. Polynomial (Duncan, 1955) was used to compare between means, using ( SAS, 2012) program .

**RESULTE AND DISCUSSION**

Table 2 present that

When measuring blood fat, it is observed from the Table 2 that there were no significant differences between the treatments in the third week of the experiment when measuring the HDL and Triglycerides, while a significant differences were found between the treatments when measuring the cholesterol and LDL were the control treatment outperformed the rest of the treatments. It is noted that there were no significant differences between treatments in the fifth week of the trial when measuring blood fat. These results are consistent with Zhu et al. (2015), when they noticed a decrease in the serum cholesterol level for the factors of adding gamma-amino butyric acid, that the lower cholesterol values for the experiment parameters may be due to the effect of the acid in the steps of making cholesterol, and the decrease in low-density lipoprotein and high-density lipoprotein are important indicators of health Birds were a high-density lipoprotein is a good cholesterol and has an important role in improving human heart health and pressure blood.

Table 2. Effect of Oral administration γ-amino butyric acid (GABA) on blood fat (average ± standard error) (average ± standard error)

|    | 3 Week | | | | 5 Week | | | |
|----|--------|---|---|---|--------|---|---|---|
|    | Cholesterol | Triglycerides | HDL | LDL | Cholesterol | Triglycerides | HDL | LDL |
| T1 | 155.71±3.20$^a$ | 54.74±16.89 | 101.93±5.99 | 23.50±2.60$^a$ | 126.70±11.42 | 86.47±9.32 | 81.33±9.80 | 21.75±8.71 |
| T2 | 106.30±7.30$^b$ | 84.03±20.56 | 116.52±9.13 | 16.00±2.08$^{ab}$ | 118.09±14.77 | 74.25±13.46 | 93.23±10.36 | 10.00±2.27 |
| T3 | 105.26±5.43$^b$ | 99.06±12.71 | 121.99±0.66 | 8.33±0.67$^c$ | 122.24±11.48 | 70.43±12.65 | 90.49±8.74 | 17.75±3.35 |
| T4 | 123.67±8.78$^b$ | 71.17±18.93 | 109.61±13.46 | 11.33±2.73$^{bc}$ | 124.55±3.50 | 74.65±10.18 | 98.90±2.22 | 10.75±0.85 |
|    | * | N.S | N.S | * | N.S | N.S | N.S | N.S |

N.S: - No significant differences between treatment.

*: - There were significant differences between the treatment at the level (P <0.05)

T1:control tratment, T2: the birds were administration0.2ml 0.4%GABA solution daily, T3: the birds were administration0.2ml 0.5%GABA solution daily and T4: the birds were administration 0.2ml 0.6%GABA solution daily

It It is noted from Table 3 that there were significant differences between treatments in the third week of the experiment when calculating blood enzymes, as T1 and T4 coefficients outperformed the rest of the treatments when calculating ALT and AST. For the fifth week of the trial, no significant differences were observed between treatments when measuring AST, while the superiority of T3 treatment was significantly higher than other treatments when measuring ALT. This result is consistent with Chen et al. (2013) where the possible explanation is that the gamma-amino butyric acid led to increase the activity of antioxidant enzymes Zhu et al. (2015).

Table 3. Effect of Oral administration γ-amino butyric acid (GABA) in blood enzymes (average ± standard error

|    | 3 Week | | 5 Week | |
|----|--------|--------|--------|--------|
|    | AST    | ALT    | AST    | ALT    |
| T1 | 225.08±7.30$^a$ | 12.99±0.05$^a$ | 311.44±29.00 | 2.66±0.10$^b$ |
| T2 | 185.49±27.38$^{ab}$ | 8.02±0.$^{93b}$ | 349.54±23.94 | 2.25±0.51$^b$ |
| T3 | 142.25±20.23$^b$ | 6.51±0.28$^b$ | 368.80±53.79 | 6.30±2.23$^a$ |
| T4 | 208.89±3.49$^a$ | 12.89±1.48$^a$ | 317.78±55.07 | 2.27±0.62$^b$ |
|    | *      | *      | N.S    | *      |

AST:- Aspartate aminotrans ferase , ALT:- Alanine aminotransferase

N.S: - No significant differences between treatment.

*: - There were significant differences between the treatment at the level (P <0.05)

T1:control tratment, T2: the birds were administration0.2ml 0.4%GABA solution daily, T3: the birds were administration0.2ml 0.5%GABA solution daily and T4: the birds were administration 0.2ml 0.6%GABA solution daily

From Figure 1, it can be noticed that there were not significant differences between treatments in the third and fifth week of experience when measuring total protein.

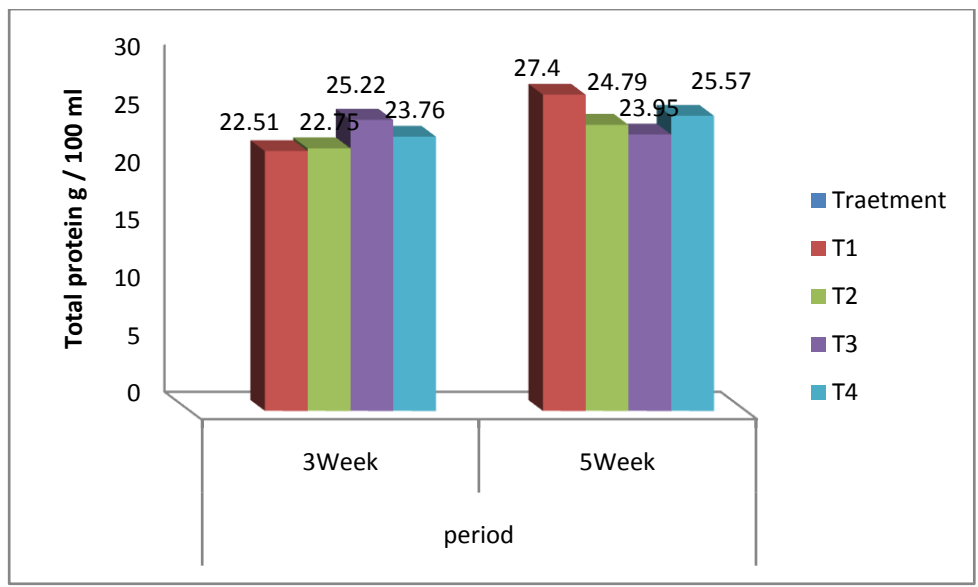

Figure 1. Effect of Oral administration γ-amino butyric acid (GABA) on total protein (average ± standard error)

From Figure 2 it is observed that there were no significant differences between the treatments when measuring glucose concentration in the blood for the third and fifth week of the experiment.

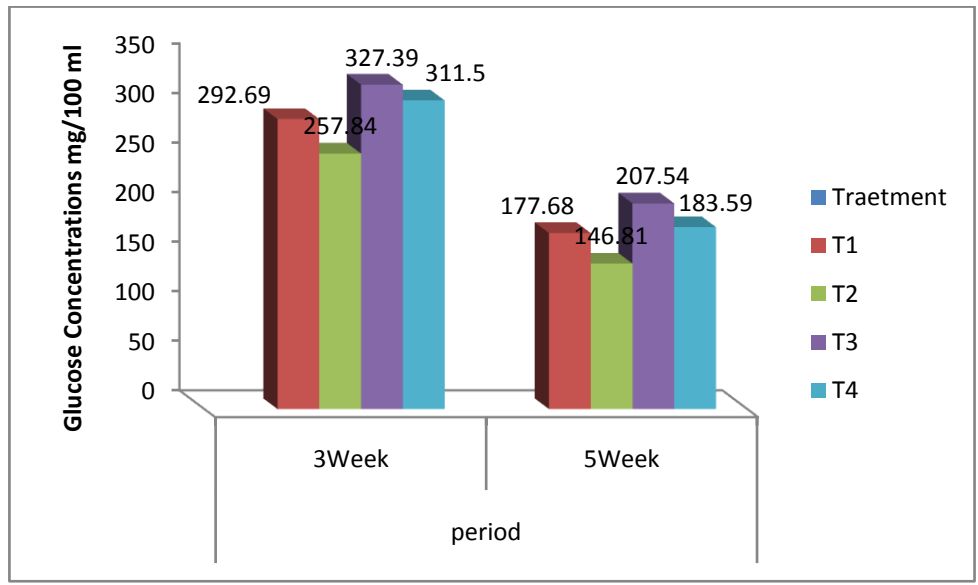

Figure 2. Effect of Oral administration γ-amino butyric acid (GABA)

in the concentration of glucose (average ± standard error)

As for measuring the concentration of calcium and phosphorous in the blood, it was noted from Figure 3,4 that there were no statistically significant differences between treatments in the third week of the experiment. However, there is significant differences found between treatments in the fifth week as administration treatments recorded a significant increase. These results are consistent with Zhu et al. (2015), that they also noticed an increase in the concentration of calcium and phosphorus in the blood. This may be due to the gamma-amino butyric acid which improving the utilization of nutrients Zhang et al (2012) through transportation and absorption of components which has a major role in regulating appetite. In addition, gamma has many antioxidant activities Huang et al. (2011), which may lead to improved production performance.

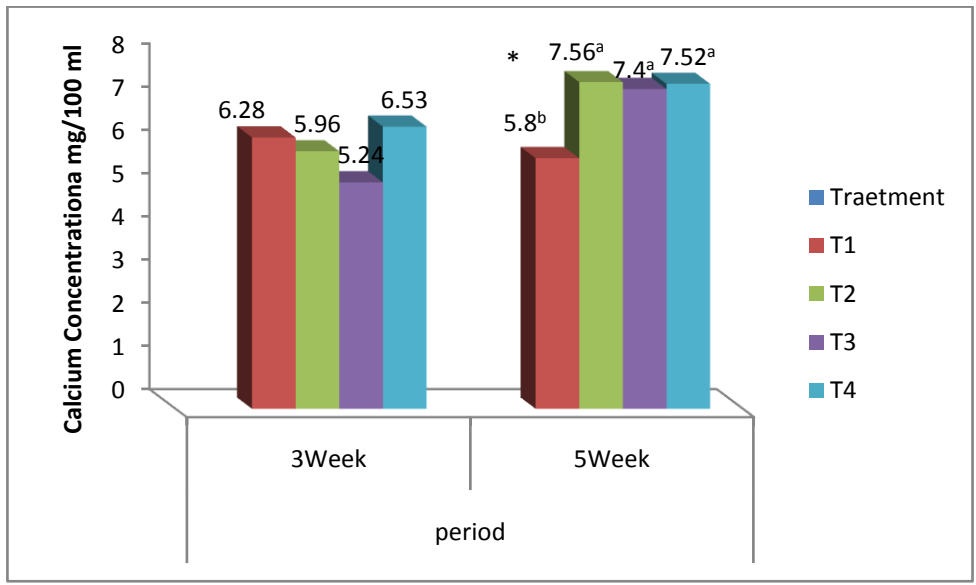

Figure 3. Effect of Oral administration γ-amino butyric acid (GABA) on the concentration of calcium (average ± standard error)

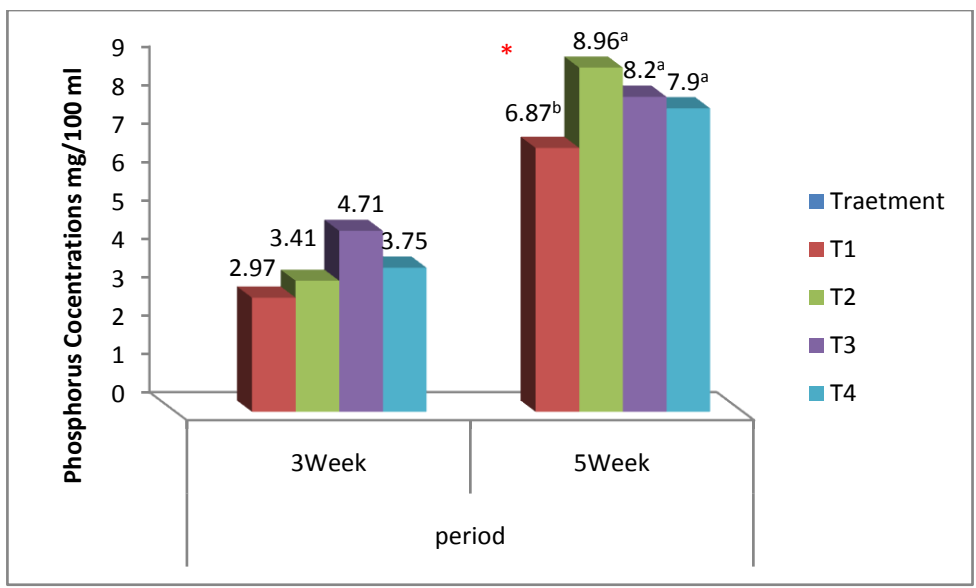

Figure 4. Effect of Oral administration γ-amino butyric acid (GABA) in the concentration of phosphorus (average ± standard error)

Figure 5 indicated that there were significant differences between the treatments when calculating the concentration of uric acid in the blood serum, as dose treatments recorded a significant decrease in the third and fifth week of the experiment. These results are not consistent with those of Zhigang et al.(2013), they noticed that there were no significant differences between the treatments when calculating the uric acid concentration. The use of gamma amino butyric acid works to preserve the heALTh of birds and organs by regulating the behavioral and physiological response Sliwowska et al. (2006), the decrease in the concentration of uric acid in the blood serum may be attributed to the improvement in the function of the kidneys, or it may be caused by a decrease in tissue demolition, or by increased utilization of protein diets.

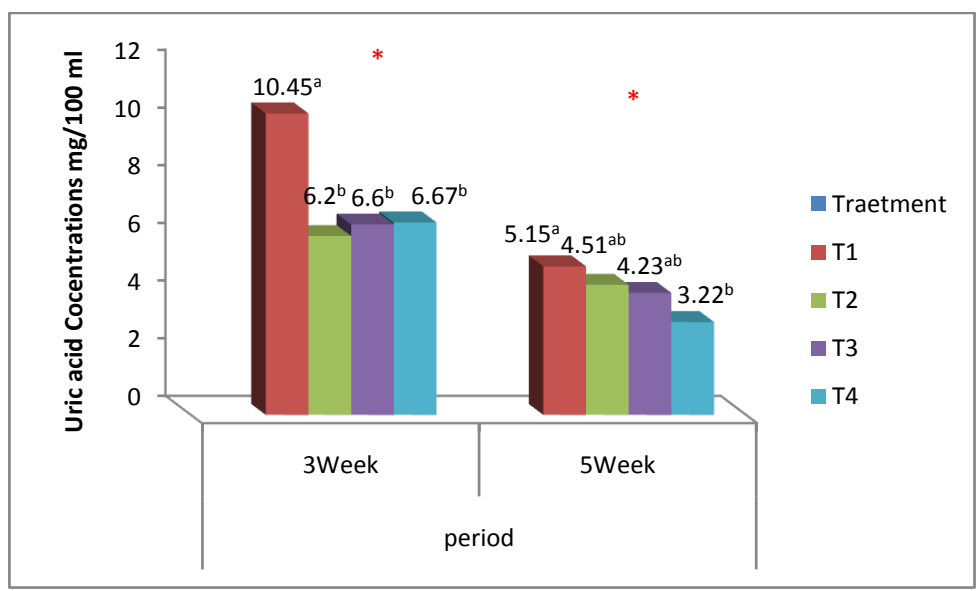

Figure 5. Effect of Oral administration γ-amino butyric acid (GABA) in the concentration of uric acid (average ± standard error)

## CONCLUSION

The addition of gamma amino butyric lion works to preserve the health of chickens by improving the transportation and absorption of mineral elements and reduces cholesterol and increases the high-density lipoprotein which considers a good cholesterol, In addition to that, it works to increase the activity of anti-oxidant enzymes. It is worthwhile to mention that the high amount of it may affects on the liver tissue and this is what we reached in this study.

## ACKNOWLEDGMENTS

The authors gratefully thank the College of Agricultural Engineering Sciences. the College of Agricultural Engineering Sciences. and the College of Veterinary Medicine

## REFERENCE

Chen Z, J. Tang, Y.Q. Sun, and J. Xie. 2013. Protective effect of γ-aminobutyric acid on anti oxidation function in intestinal mucosa of Wenchang chicken induced by heat stress. Journal of Animal and Plant Sciences 23, 1634–1641.

Chen Z. , J. Xie , B. Wang , and J. Tang. 2014. Effect of γ-amino butyric acid on digestive enzymes, absorption function, and immune function of intestinal mucosa in heat-stressed chicken. 2014 Poultry Science 93 :2490–2500

Dai, S.F; F. Gao, W.H. Zhang, S.X. Song, X.L. Xu, and G.H. Zhou. 2011. Effects of dietary glutamine and gamma-aminobutyric acid on performance, carcass characteristics and serum parameters in broilers under circular heat stress. Anim. Feed Sci. Tech., 168: 51-60.

Duncan, D. B. 1955. Multiple range and multiple F tests. Biometrics. 11:1–42.

Hashimoto, K. 2011. The role of glutamate on the action of antidepressants. Prog. Neuro-Psychop.35, p.1558-1568.

Hayakawa, K., M. Kimura, K. Kasaha, K. Matsumoto, H. Sansawa, and Y. Yamori. 2004. Effect of a γ-aminobutyric acid-enriched dairy product on the blood pressure of spontaneously hypertensive and normotensive Wistar-Kyoto rats. Brit. J. Nutr. 92,p.411-417.

Huang H.L., W.J. Zhao, X.T. Zou, H. Li, M. Zhang, and X.Y. Dong. 2011. Effect of γ-aminobutyric acid on incubation, immunity and antioxidation activity in pigeon. Chin J. Vet. Sci. 9(31): 1327-1331.

Jonaidi H., L. Abbassi, M.M. Yaghoobi, H. Kaiya, D.M. Denbow, Y. Gammali, and B. Shojaei. 2012.The role of GABAergic system on the inhibitory effect of ghrelin on food intake in neonatal chicks. Neurosci. Letters. 520(1): 82-86.

Li, H. , T. Qiu, G. Huang, and Y. Cao. 2010. Production of gamma-aminobutyric acid by *Lactobacillus brevis* NCL912 using fed-batch fermentation. Microb Cell Fact . 9: 85. doi: 10.1186/1475-2859-9-85

NRC., 1994. National Research Council: Nutrient Requairment for Poultry. 9th Rev. Edn., National Academy Press, USA.

Oh S.H, and W.G. Choi. 2000. Production of the quality germinated brown rices containing high γ-aminobutyric acid by chitosan application. Korean Journal of Biotechnology and Bioengineering 15, 615–620.

Omori M., T. Yano, J. Okamoto, T. Tsushida, T. Murai, M. Higuchi .1987. Effect of anaerobically treated tea (gabaron tea) on blood pressure of spontaneously hypertensive rats. Nippon Nogeikagaku Kaishi 61, 1449–1551.

Park J.H. and I.H. Kim. 2015. Effects of dietary gamma-aminobutyric acid on egg production, egg quality, and blood profiles in layer hens. Veterinarni Medicina, 60, 2015 (11): 629–634.

SAS. 2012. SAS/State User's Guide. SAS Inst. Inc., Cary, NC.

Shekhar A, P.L. Johnson, T.J. Sajdyk, S.D. Fitz, S.R. Keim, P.E. Kelley, D.K. Smith, T. Kassam, B. Singh, and J.F. Elliott.1992. Escherichia coli has two homologous glutamate decarboxylase genes that map to distinct loci. Journal of Bacteriology 174, 5820–5826.

Sliwowska, J. H., H. J. Billings, R. L. Goodman, and M. N. Lehman. 2006. Immunocytochemical colocalization of GABA-B receptor subunits in gonadotropin-releasing hormone neurons of the sheep. Neuroscience 141:311-319.

Wang, Y, Y. Xian-Liang, C. Yuan, C. Lie-Ran, and X. Xue-Ping .2011. Effects on growth of chickens breAST muscle by adding γ-aminobutyric acid (gaba) and vitamin E (VE) to diets. J. Anim. Sci. Technol. Uni., 25: 5-9.

Zhang M., X.T. Zou, H. Li, X.Y. Dong, W. Zhao. 2012. Effect of dietary γ-aminobutyric acid on laying performance, egg quality, immune activity and endocrine hormone in heat-stressed Roman hens. Animal Science Journal 83, 141–147.

Zhigang, S., A. Sheikhahmadi, and Z. Li. 2013. Effect of dietary γ-aminobutyric acid on performance parameters and some plasma metabolites in Cherry Valley ducks under high ambient temperature. Iranian Journal of Veterinary Research, Shiraz University, 2013, Vol. 14, No. 4, Pages 283-290.

Zhu Y.Z., J.L. Cheng, M. Ren, L. Yin, X.S. Piao. 2015. Effect of γ-aminobutyric acid-producing Lactobacillus strain on laying performance, egg quality and serum enzyme activity in Hy-Line brown hens under heat stress. Asian Australasian Journal of Animal Sciences 28, 1006–1013.